\documentclass{article}

\usepackage{latexsym}
\usepackage{amssymb}
\usepackage{amsmath}
\numberwithin{equation}{section}

\newcommand{\beq}{\begin{equation}}
\newcommand{\eeq}{\end{equation}}
\newcommand{\bea}{\begin{eqnarray}}
\newcommand{\eea}{\end{eqnarray}}
\newcommand{\barr}[1]{\begin{array}}
\newcommand{\earr}{\end{array}}
\newtheorem{theorem}{Theorem}[section]
\newtheorem{definition}{Definition}[section]

\newcommand{\bdf}{\begin{definition}}
\newcommand{\edf}{\end{definition}}
\newcommand{\bth}{\begin{theorem}}
\newcommand{\enth}{\end{theorem}}
\newcommand{\pd}{\partial}

\def\c+{\rlap{\ \raisebox{.2ex}{\scriptsize+}}\supset}
\begin{document}

\title{Contact transformations for difference schemes}
\author{ Decio Levi~$^1$, Christian Scimiterna~$^2$, Zora Thomova~$^3$ and Pavel Winternitz~$^4$}
\maketitle
\noindent
{$^1$ Dipartimento di Ingegneria Elettronica, Universit\`a degli Studi Roma Tre and Sezione INFN, Roma Tre, Via della Vasca Navale, 84, 00146 Roma, Italy \\
$^2$ Dipartimento di Fisica and Ingegneria Elettronica, Universit\`a degli Studi Roma Tre and Sezione INFN, Roma Tre, Via della Vasca Navale, 84, 00146 Roma, Italy \\
$^3$ Department of Engineering, Sciences and Mathematics, SUNY Institute of Technology, 100 Seymour Road, Utica, NY 13502, USA\\
$^4$ Centre de recherche math{\'e}matiques and D{\'e}partement de math{\'e}matiques et de statistique,
Universit{\'e} de Montr{\'e}al, Case postale 6128, succ. centre-ville, Montr{\'e}al, Qu{\'e}bec, H3C 3J7, Canada}

\begin{abstract}
We define a class of transformations of the dependent and independent variables in an ordinary difference scheme.
The transformations leave the solution set of the system invariant and reduces to a group of contact transformations
in the continuous limit. We use a simple example to show that the class is not empty and that such ``contact
transformations for discrete systems'' genuinely exist.
\end{abstract}

\section{Introduction}
In a recent article \cite{LTW1} we proposed a definition of a contact transformation
for an ordinary difference scheme ($O\Delta S$) of order $K$
\bea
E_a ( \{ x_k \}, \{y_k \}, k=n+M, n+M+1, \ldots , n+N) =0 \label{e1-1} \\
a=1,2 \quad   K=N-M+1, \quad  n,M,N \in  \mathbf{Z}, \quad N>M. \nonumber
\eea
The Lie algebra of the symmetry group of contact transformations in \cite{LTW1} was realized by vector fields of the form
\bea
X= \xi_n \pd_{x_n}+\phi_n \pd_{y_n}  \label{e1-2}
\eea
where the functions $\xi_n$ and $\phi_n$ satisfy the following conditions:
\begin{itemize}
\item
They depend on $J \in \mathbf{Z}^+$ points $(x_{n+k}, y_{n+k}), k=0,1,\ldots J-1$ with $J\geq 2$ for at least one vector
field in the symmetry algebra;
\item
The Lie algebra should be integrable to a Lie group. This implies that all coefficients in the $J-th$ prolongation of
the vector fields $X$ should depend only on $(x_{n+k}, y_{n+k}), 0\leq k \leq J $, and not on any further points;
\item
In the continuous limit the algebra and the symmetry group of the
$O\Delta S$ (\ref{e1-1}) should reduce to the Lie algebra and Lie group of the ordinary differential equation (ODE)
that is the continuous limit of (\ref{e1-1}).
\end{itemize}

It was shown in \cite{LTW1} that contact transformations satisfying theses conditions do not exist. The theorems proven in \cite{LTW1} are difference
analogues of B{\"a}cklund's famous theorem \cite{Backlund} from which it follows that contact transformations can only depend on the first derivatives. Indeed, those that depend on higher derivatives are prolongations of point transformations \cite{Backlund,Ibr}.

The negative result presented in \cite{LTW1} leaves open the possibility that a more general class of transformations exists,
 that takes solutions of the $O \Delta S$ (\ref{e1-1}) into solutions, and reduces to contact transformations in the continuous limit. The purpose of this article is to show that this is indeed the case.

In Section 2 we consider a general $n-th$ order ODE
and replace it by a system of two lower order equations. We show that some point symmetries of the system give rise
to contact transformations for the original n-th order equation. In particular we show that the point symmetries
of the system $u'=v, v''=0$ give rise to the 10 dimensional invariance group of contact transformations of the ODE $u'''=0$.

In Section 3 we apply the same approach to difference systems. We discretize system $u'=v, v''=0$ in a manner that
preserves a 7-dimensional subalgebra of its Lie point symmetry algebra. We then eliminate the variable $v$ from the system and obtain a third order $O\Delta S$ for the variable $u(x)$ allowing a group of transformations that in continuous limit include contact transformations.

Section 4 is devoted to conclusions.  We suggest there a less restrictive definition of contact transformations for
$O \Delta S$ then the one given in \cite{LTW1}.
\section{Contact symmetries of an ODE as a point symmetries of a system of ODEs}
Let us consider an m-th order ODE
\beq
u^{(m)} = F(x, u, u', \ldots, u^{(m-1)}), \, \, \, \, \, m \geq 3. \label{2-1}
\eeq
It can always be replaced by a system of lower order equations, for instance by putting
\beq
u'=v, \, \, \, \ \, \, \, v^{(m-1)}=F(x, u, v, v', \ldots, v^{(m-2)}). \label{2-2}
\eeq
With this choice contact symmetries of (\ref{2-1}) are reduced to point symmetries of the system (\ref{2-2}).
The Lie algebra of {\it point symmetries} of the system (\ref{2-2}) is realized by vector fields of the form
\beq
X=\xi(x,u,v) \pd_x +\phi(x,u,v) \pd_u +\psi(x,u,v) \pd_v \label{e2-3}
\eeq
such that the prolongation $pr^{(n-1)} X$ of $X$ annihilates the system (\ref{2-2}) on its solution set.

Since the solution sets of the single equation (\ref{2-1}) and the system (\ref{2-2}) coincide the vector fields (\ref{e2-3}) will also generate symmetry transformations for the ODE (\ref{2-1}).
Returning to the original variables in (\ref{e2-3}) we have
\beq
X=\xi(x,u,u')\pd_x +\phi(x,u,u') \pd_u + \phi^{(1)}(x,u,u') \pd_{u'}.\label{2-4}
\eeq
Thus, if $\xi$ or $\phi$ in (\ref{e2-3}) depend on $v$ then (\ref{2-4}) will correspond to
the first prolongation of a contact transformation for the ODE (\ref{2-1}) with
\beq
\psi(x,u,v)=\phi^{(1)}(x,u,u'). \label{e2-5}
\eeq
Proceeding in this manner we find all point symmetries of the system (\ref{2-2})
and some, if not all
contact symmetries of the ODE (\ref{2-1}).

Let us now consider the converse problem. Given a finite dimensional Lie algebra of vector fields of the form (\ref{e2-3}),
what is the most general system of two second order equations of the form (\ref{2-2}), invariant under the corresponding group of  point transformations? For simplicity we concentrate on the case of a second order system, i.e. $n=3$ in (\ref{2-1}) and (\ref{2-2}).

We choose a basis $\{X_1, \ldots, X_s \}$ for the given Lie algebra and write the second order prolongations of the basis vectors as
\bea
pr^{(2)}X_i&=&\xi_i \pd_x +\phi_i \pd_u +\psi_i \pd_v+\phi^x \pd_{u'}+
\psi^x \pd_{v'} +\phi^{xx} \pd_{u''}+\psi^{xx} \pd_{v''}, \label{e2-6}\\
i&=&1, \ldots, s. \nonumber
\eea
To find the differential invariants we must solve the system of quasilinear PDEs
\beq
pr^{(2)}X_i F(x,u,v,u',v',u'',v'')=0, \, \, \, \, \, \, \,  i=1, \ldots, s. \label{2-7}
\eeq
The system can be written as a system of equations in matrix form as
\beq
MU =0, \, \, \, \, \, \, \, \, \, \, U^T =(F_x, F_u, F_v, F'_{u},F'_{v},F''_{u},F''_{v}). \label{e2-7}
\eeq

``Strong invariants'' are obtained if the matrix $M$ has maximal rank $r_M=min(N,7)$. The number of such invariants is
$N_S=7-r_M$ and they are invariant on the entire jet space.

The rank of $M$ can be $r < r_M$ on some submanifold and then further invariants, ``weak invariants'' can exist. The number of all invariants is  $N=7-r$.

Let us now consider a specific example for which all contact symmetries are obtained from the point symmetries of a lower order system.
The equation (\ref{2-1}) is specified to
\beq
u'''=0 \label{2-9}
\eeq
and the system (\ref{2-2}) is
\beq
u'=v, \, \, \, \, \, \, \, \, \, \,  v''=0. \label{2-10}
\eeq
Following a standard procedure \cite{olver} we find that the Lie point symmetry algebra of (\ref{2-10}) is isomorphic to the de Sitter algebra $o(3,2)$ realized by vector fields with a basis:
\bea
X_1&=&\pd_u  \nonumber \\
X_2&=&x \pd_u +\pd_v \nonumber \\
X_3&=& \pd_x \nonumber \\
X_4&=&x^2 \pd_u +2x \pd_v \nonumber \\
X_5&=&x\pd_x -v \pd_v \label{2-11}\\
X_6&=&u \pd_u + v \pd_v \nonumber \\
X_7&=&2v \pd_x +v^2 \pd_u \nonumber \\
X_8&=&x^2 \pd_x +2xu \pd_u +2u \pd_v \nonumber \\
X_9&=&2(xv-u) \pd_x +xv^2 \pd_u +v^2 \pd_v \nonumber \\
X_{10}&=& 2x (2u-xv) \pd_x +(4 u^2 -x^2 v^2) \pd_u +2v (2u-xv) \pd_v.  \nonumber
\eea
We shall denote the corresponding local Lie group $G$.

By construction (\ref{2-11}) is the Lie algebra of {\it point} transformations of the {\it system} (\ref{2-10}). If we eliminate $v$ from (\ref{2-10}) we return to the {\it single} third order ODE (\ref{2-9}). Eliminating $v$ in the same way from the vector field (\ref{2-11}) we obtain the {\it first prolongation} of the Lie algebra of {\it contact} symmetries of (\ref{2-9}) in the form (\ref{2-4}).

Specifically for the vector fields (\ref{2-11}), we see that $\xi$ and $\phi$ do not depend on $v$ for $X_1, \ldots, X_6 $ and $X_8$ so they will generate point symmetries for (\ref{2-1}). On the other hand $X_7, X_9$ and $X_{10}$ turn into contact symmetries of (\ref{2-1}) after the substitution (\ref{e2-5}).

We mention that this is an alternative way of calculating the group of contact symmetries for (\ref{2-9}) to the standard one, used for example in Ref \cite{hydon}. The result is of course same.

Let us now consider the converse problem for the subalgebra $L_0=\{ X_1, \ldots, X_7 \}$ where $X_7$ generates a contact transformation for the ODE (\ref{2-9}).

The matrix $M$ of (\ref{e2-7}) has $r_M=7$ so the group corresponding to $L_0$ has no strong invariants.

However for $u'=v, \, \, v''=0$ we have $r(M)=5$ and on this manifold the system of ODEs (\ref{2-10}) is invariant (or at least ``weakly invariant''). Moreover, applying the prolongations of $X_8, X_9$ and $X_{10}$ to the system (\ref{2-2}) we verify that this system is invariant under the entire de Sitter group $G \sim O(3,2)$. To sum up, the only invariants of $G \sim O(3,2)$ and $G_0$ in this case are weak ones, namely
\beq
I_1 = u'- v =0, \qquad  {\rm and} \qquad I_2=v''=0. \label{e2-11a}
\eeq

All subgroups of $SO(3,2)$ were classified into conjugacy classes in Ref \cite{PSW77}. The subgroup $G_0 \subset G$ corresponding to the Lie algebra
$L_0 = \{X_1, \ldots, X_7\}$ is isomorphic to one of the 7 maximal subgroups of $SO(3,2)$ called the ``optical group'' $Opt(2,1)$ of 2+1 dimensional Minkowski space. The algebra $L_0$ was already known to S. Lie \cite{Lie74, Lie90}
as was its 6 dimensional subalgebra $\tilde{L_0}=\{X_1, X_2, X_3, X_4, X_5,X_7 \}$. The algebras $L, L_0, \tilde{L_0}$ are the only finite dimensional
Lie algebras of contact transformations of a complex plane (other than point transformations) \cite{Lie74, Lie90}.


\section{Contact symmetries of an $O \Delta S $ as point symmetries of a lower order system?}

Let us now try to produce an analogue of contact transformations for a difference system in the same manner as we did for the ODE (\ref{2-9}) in Section 2.

We again start from the subalgebra $L_0=\{X_1, \ldots X_7 \}$ and look for a difference system that allows $L_0$ as
a Lie point symmetry algebra. We use the formalism for symmetries of $O \Delta S$ as outlined in \cite{LTW1, lw2,doro2}.
The idea is to construct a 3 point difference system
\beq
E_a(\{x_k,u_k,v_k\}_{k=n,n+1,n+2})=0, \, \, \, \, a=1,2,3 \label{e3-1}
\eeq
which has $L_0$ as its symmetry algebra and reduces to the system (\ref{2-10}) in the continuous limit.

In order to facilitate the continuous limit we use the variables
\bea
x_n, & h_{n+1}=x_{n+1}-x_n, & h_{n+2}=x_{n+2}-x_{n+1} \nonumber \\
u_n, & p_{n+1}^{(1)}=\frac{u_{n+1}-u_n}{x_{n+1}-x_n}, &
p_{n+2}^{(2)}=2 \frac{p^{(1)}_{n+2}-p^{(1)}_{n+1}}{x_{n+2}-x_n} \label{e3-2} \\
v_n, & q_{n+1}^{(1)}=\frac{v_{n+1}-v_n}{x_{n+1}-x_n}, &
q_{n+2}^{(2)}=2 \frac{q^{(1)}_{n+2}-q^{(1)}_{n+1}}{x_{n+2}-x_n} \nonumber
\eea
instead of ${x_{n+k},y_{n+k}}, k=0,1,2$.

The relevant prolongations of the vector fields in the variables have the form:
\bea
pr X&=&\xi_n \pd_{x_n} +\phi_n \pd_{u_n}+\psi_n \pd_{v_n} +
\phi^{(1)}_{n+1} \pd_{p^{(1)}_{n+1}}+\psi^{(1)}_{n+1} \pd_{q^{(1)}_{n+1}} \nonumber \\
&+&\phi^{(2)}_{n+2} \pd_{p^{(2)}_{n+2}}+
\psi^{(2)}_{n+2} \pd_{q^{(2)}_{n+2}}
+\lambda^{(1)} \pd_{h_{n+1}} +\lambda^{(2)} \pd_{h_{n+2}}.\nonumber
\eea
Here $\lambda^{(k)}, \phi^{(k)}$ and $\psi^{(k)}$ are calculated using the results given in I, namely
\bea
&& \lambda^{(k)} = \xi_{n+k} - \xi_{n+k-1} \label{e3-4} \\
&& \phi^{(k)}_{n+k}=\frac{k}{\sum_{j=1}^k h_{n+j}} h_{n+k} \Delta^T
\phi^{(k-1)}_{n+k-1}-p^{(k)}_{n+k} \frac{1}{\sum_{j=1}^k h_{n+j}}
\sum_{j=1}^{k} h_{n+j} \Delta^T \xi_{n+j-1} \nonumber \\
&& \psi^{(k)}_{n+k}=\frac{k}{\sum_{j=1}^k h_{n+j}} h_{n+k} \Delta^T
\psi^{(k-1)}_{n+k-1}-q^{(k)}_{n+k} \frac{1}{\sum_{j=1}^k h_{n+j}}
\sum_{j=1}^{k} h_{n+j} \Delta^T \xi_{n+j-1} \nonumber
\eea
where $\Delta^T$ is the total difference operator
\bea \nonumber
\Delta^T F(x_n,u_n,v_n,p^{(1)}_{n+1},q^{(1)}_{n+1},...)&=&
\frac{1}{h_{n+1}}  \{ F(x_{n+1},u_{n+1},v_{n+1},p^{(1)}_{n+2},q^{(1)}_{n+2},...) \nonumber \\
&&-F(x_n,u_n,v_n,p^{(1)}_{n+1},q^{(1)}_{n+1},...) \}.
\eea
More specifically we have
\bea
pr X_1&=&\pd_{u_n} \label{e3-6}\\
pr X_2 &=&x_n \pd_{u_n} +\pd_{v_n}+\pd_{p^{(1)}_{n+1}} \nonumber \\
pr X_3 &=& \pd_{x_n} \nonumber \nonumber \\
pr X_4 &=& x^2_n \pd_{u_n} +2 x_n \pd_{v_n} +(2x_n+h_{n+1}) \pd_{p^{(1)}_{n+1}} +2 \pd_{q^{(1)}_{n+1}} +2 \pd_{p^{(2)}_{n+2}} \nonumber \\
prX_5 &=& x_n \pd_{x_n} -v_n \pd_{v_n} -p^{(1)}_{n+1} \pd_{p^{(1)}_{n+1}}-2q^{(1)}_{n+1} \pd_{q^{(1)}_{n+1}}-
2 p^{(2)}_{n+2} \pd_{p^{2)}_{n+2}} -
3 q^{(2)}_{n+2} \pd_{q^{(2)}_{n+2}}\nonumber \\
&& +h_{n+1} \pd_{h_{n+1}} +
h_{n+2} \pd_{h_{n+2}} \nonumber \\
pr X_6 &=& u_n \pd_{u_n} +  v_n \pd_{v_n} +p_{n+1}^{(1)} \pd_{p_{n+1}^{(1)}}
+ q_{n+1}^{(1)} \pd_{q_{n+1}^{(1)}} +p_{n+2}^{(2)} \pd_{p_{n+2}^{(2)}}
+q_{n+2}^{(2)} \pd_{q_{n+2}^{(2)}} \nonumber   \\
pr X_7 &=& 2 v_n \pd_{x_n} +v_n^2 \pd_{u_n}
+2 h_{n+1} q^{(1)}_{n+1} \pd_{h_{n+1}} \nonumber \\
&& + 2 h_{n+2}( q^{(1)}_{n+1} + \frac{h_{n+1}+h_{n+2}}{2} q^{(2)}_{n+2})\pd_{h_{n+2}} \nonumber  \\
&&+(2 v_n q^{(1)}_{n+1} -2 p^{(1)}_{n+1} q^{(1)}_{n+1}+h_{n+1}  (q^{(1)}_{n+1})^2) \pd_{p^{(1)}_{n+1}} \nonumber \\
&& -2 (q_{n+1}^{(1)})^2 \pd_{q_{n+1}^{(1)}} \nonumber \\
&& + \phi_{n+2}^{(2)} \pd_{p_{n+2}^{(2)}} + \psi_{n+2}^{(2)} \pd_{q_{n+2}^{(2)}} \nonumber
\eea
where
\bea
\phi_{n+2}^{(2)} &=&
2 q_{n+1}^{(1)} q_{n+2}^{(2)} h_{n+1} - h_{n+1} p_{n+2}^{(2)} q_{n+2}^{(2)}
+ \frac{1}{2} (q_{n+2}^{(2)})^2 (h_{n+2})^2 +2 (q_{n+1}^{(1)})^2 \nonumber \\
&& + 2 v_n q_{n+2}^{(2)} + 2 q_{n+1}^{(1)} q_{n+2}^{(2)} h_{n+2} -2 h_{n+2} p_{n+2}^{(2)} q_{n+2}^{(2)} \nonumber \\
&& +\frac{1}{2} (q_{n+2}^{(2)})^2 h_{n+2} h_{n+1} \nonumber \\
&& - 4 p_{n+2}^{(2)} q_{n+1}^{(1)} - 2 p_{n+1}^{(1)} q_{n+2}^{(2)} \nonumber \\
\psi_{n+2}^{(2)}&=& -h_{n+1} (q_{n+2}^{(2)})^2 - 2 h_{n+2} (q_{n+2}^{(2)})^2 \nonumber
- 6 q_{n+1}^{(1)} q_{n+2}^{(2)}.
\eea
To calculate invariants of the subgroup corresponding to (\ref{e3-6}) we
impose
\bea
&&pr X_a F(x_n,u_n,v_n,p^{(1)}_{n+1},q^{(1)}_{n+1},p^{(2)}_{n+2},q^{(2)}_{n+2},
h_{n+1},h_{n+2})=0, \label{e3-7} \\
&& \qquad \qquad a=1, \ldots, 7. \nonumber
\eea
In matrix form (\ref{e3-7}) can be written as
\beq \nonumber
MU_d=0, \, \, \, \, \, \, \, \, \, U^T_d=(F_{x_n}, F_{u_n}, F_{v_n}, F_{p^{(1)}_{n+1}},F_{q^{(1)}_{n+1}},F_{p^{(2)}_{n+2}},
F_{q^{(2)}_{n+2}},F_{h_{n+1}},F_{h_{n+2}}).
\eeq
As in the continuous case, to get a sufficient number of invariants we must restrict to an invariant manifold on which the rank of $M$ is $r(M)=5$, rather than $r(M)=7$ as in the generic case. The invariant manifold in this case is given by
\bea
I_1=p^{(1)}_{n+1}-v_n-\frac{1}{2} q^{(1)}_{n+1}=0, \, \, \, \, \, \, \, I_2=q^{(2)}_{n+2}=0. \label{e3-9}
\eea
and on this manifold the invariants are
\bea
&&I_1=0, \, \, \, \, \, I_2=0 \nonumber \\
&&I_3=\frac{h_{n+2}}{h_{n+1}}, \, \, \, \, \, \, I_4=(h_{n+1})^2
(p^{(2)}_{n+2}-q^{(1)}_{n+1}) \label{e3-10}
\eea
An invariant difference scheme that reduces to the system (\ref{2-2}) in the continuous limit $h_{n+k}\rightarrow 0$ is
\beq
p_{n+1}^{(1)}-v_n-\frac{1}{2}h_{n+1}q^{(1)}_{n+1}=0, \, \, \, \, \, \, q_{n+2}^{(2)}=0, \, \, \, \, \, \,  h_{n+2}=c h_{n+1} \label{e3-11}
\eeq
where $c$ is an arbitrary real constant and in particular $c=1$ corresponds to a uniform lattice.

The $O \Delta S$ (\ref{e3-11}) is invariant under the group $G_0$.

Let us now eliminate the variable $v_n$ from the system (\ref{e3-11}).
Taking the discrete derivative of the first equation in (\ref{e3-11}) and using the second equation $q^{(2)}_{n+2}=0$ we obtain
\beq
q^{(1)}_{n+1}=p^{(2)}_{n+2} \label{e3-12}
\eeq
thus $I_1=I_2=0$ implies $I_4=0$ in (\ref{e3-10}). Taking the discrete derivative of \ref{e3-12} we obtain
\beq
q^{(2)}_{n+2}=\frac{2(h_{n+1}+h_{n+2}+h_{n+3})}{3(h_{n+1}+h_{n+2})} p^{(3)}_{n+3} \label{e3-13}
\eeq
and hence we have
\beq
p^{(3)}_{n+3}=0, \, \, \, \, \, h_{n+2}=ch_{n+1} \label{e3-14}
\eeq
as a consequence of (\ref{e3-11}).

The difference system (\ref{e3-11}), together with its difference consequence (\ref{e3-12}) is thus invariant under the Lie group $G_0$. Moreover, the equations $I_1=I_2=I_4=0$ are invariant under the entire group $G \sim O(3,2)$ however $I_3=c$ is invariant only under $G_0 \sim Opt(2,1)$. For the system (\ref{e3-11}) $G$ and $G_0$ are groups of point transformations.

Now let us consider the third order difference scheme
\beq
p^{(3)}_{n+3}=0, \, \, \, \, \, \, \, h_{n+2}=c h_{n+1}. \label{e3-14-2}
\eeq
We can obtain its symmetry algebra from $L_0$ by eliminating $v_n, q^{(1)}_{n+1}$ and $q^{(2)}_{n+2}$ from all the expressions in (\ref{e3-6}). From (\ref{e3-11}) and (\ref{e3-12}) we have
\beq
v_n=p_{n+1}^{(1)}-\frac{1}{2} h_{n+1}p^{(2)}_{n+2}. \label{e3-15}
\eeq
To get the actual vector fields of the symmetry algebra of the $O \Delta S$ (\ref{e3-14}) (as opposed to their prolongations) we need to keep only the coefficients of $\pd_{x_n}$ and $\pd_{u_n}$. From (\ref{e3-6}) we see that $ \{ X_1, \ldots, X_6 \}$ remain as point transformations for the $O \Delta S$, however $X_7$ corresponds to a contact transformation
\beq
X_7=\left( p^{(1)}_{n+1} - \frac{1}{2} h_{n+1} p^{(2)}_{n+2} \right) \pd_{x_n}
+ \left( p^{(1)}_{n+1} - \frac{1}{2} h_{n+1} p^{(2)}_{n+2}\right)^2 \pd_{u_n}. \label{e3-16}
\eeq
The third prolongation of $X_7$ will have the form
\beq
pr X_7 = X_7 + \phi^{(1)} \pd_{p^{(1)}_{n+1}} + \phi^{(2)} \pd_{p^{(2)}_{n+2}}+\phi^{(3)} \pd_{p^{(3)}_{n+3}} +\lambda^{(1)} \pd_{h_{n+1}} + \lambda^{(2)} \pd_{h_{n+2}}+\lambda^{(3)} \pd_{h_{n+3}}
\label{e3-17}
\eeq

The coefficients $\phi^{(k)}, \lambda^{(k)}$ in (\ref{e3-17}) were calculated using Maple and they are too long to reproduce here. The important fact is that we have
\bea
\phi^{(k)}&=&\phi^{(k)} (p_{n+j+2}^{(j+2)},h_{n+j+2}), \qquad \qquad 0\leq j\leq k \label{e3-19} \\
\lambda^{(k)}&=&\lambda^{(k)}(p_{n+j+2}^{(j)},h_{n+j+2}).
\eea
Thus the algebra cannot be integrated to Lie group as shown in general in our previous article \cite{LTW1}.

However, on the solution set of the $O\Delta S$ (\ref{e3-14-2}) we have $p_{n+3}^{(3)}=0$ and $pr^{(3)}X_7$ (\ref{e3-17}) simplifies to
\bea
prX_7|_{p_{n+3}^{(3)}=0}&=&(p_{n+1}^{(1)}-\frac{1}{2}h_{n+1}p_{n+2}^{(2)})\pd_{x_n}+
(p_{n+1}^{(1)}-\frac{1}{2}h_{n+1}p_{n+2}^{(2)})^2 \pd_{u_n} \nonumber \\
&& +0 \, \pd_{p_{n+1}^{(1)}}-(p_{n+2}^{(2)})^2 \pd_{p_{n+2}^{(2)}}+0 \, \pd_{p_{n+3}^{(3)}} \label{e3-19b} \\
&&+p_{n+2}^{(2)}(h_{n+1}\pd_{h_{n+1}}+h_{n+2}\pd_{h_{n+2}}+h_{n+3}\pd_{h_{n+3}}). \nonumber
\eea
This can be integrated to give a one parameter group of contact transformations, namely
\bea
\tilde{x}_n &=&x_n+\lambda \left( p_{n+1}^{(1)}-\frac{1}{2}h_{n+1}p_{n+2}^{(2)} \right) \nonumber \\
\tilde{u}_n &=&u_n+\frac{1}{2}\lambda \left( p_{n+1}^{(1)}-\frac{1}{2}h_{n+1}p_{n+2}^{(2)} \right)^2 \label{e3-20} \\
\tilde{p}_{n+1}^{(1)}&=& p_{n+1}^{(1)}, \qquad \tilde{p}_{n+2}^{(2)}=\frac{p_{n+2}^{(2)}}{1+\lambda p_{n+2}^{(2)}}, \qquad \tilde{p}_{n+3}^{(3)}=p_{n+3}^{(3)} \nonumber \\
\tilde{h}_{n+k}^{(k)}& = &h_{n+k} (1+\lambda p_{n+2}^{(2)}), \qquad k=1,2,3.  \nonumber
\eea
In the continuous case we obtain
\beq
\tilde{x}=x+\lambda u_x, \qquad \tilde{u}=u+ \frac{1}{2}\lambda (u_x)^2, \qquad
\tilde{u}_{\tilde{x}}=u_x. \label{e3-21}
\eeq
It is easy to see that the (\ref{e3-21}) leaves the ODE (\ref{2-9}) invariant and the $ O \Delta S$ (\ref{e3-14-2}) is invariant under the transformation (\ref{e3-20})




\section{Conclusions}

The main conclusion is that the definition of contact symmetries for difference schemes given in \cite{LTW1} was too restrictive.
It required that the Lie algebra of contact transformations be integrable to a Lie group on the entire jet space.

Let us propose a less restrictive definition
\bdf
The vector fields (\ref{e1-2}) where $\xi_n$ and $\phi_n$ are
functions of \\
\noindent
$\{ x_{n+j}, y_{n+j}, j=1, \ldots J \}$ form a Lie algebra of contact symmetries of $O \Delta S$ (\ref{e1-1}) if they satisfy the following conditions
\begin{itemize}
\item
\beq
pr^{(N)} X E_a|_{E_1=E_2} =0, \, \, \, a=1,2;
\eeq
\item
At least one of the vector fields in the Lie algebra has $J \geq 2$;
\item
The Lie algebra should be integrable to a Lie group {\it at least on the invariant surface defined by the $ O \Delta S$ (\ref{e1-1}});
\item
In the continuous limit the symmetry algebra and Lie group reduce to a Lie algebra and group of contact symmetries of the corresponding ODE.

\end{itemize}
\edf

The algebra (\ref{e3-6}) constructed in Section 3 is the algebra of contact symmetries of the $O \Delta S$ (\ref{e3-14}). This provides an example of the fact that Definition 1 is not empty: such contact symmetries of $ O \Delta S$ do exist!

Some more specific conclusions concerning the example of ODE $y'''=0$ can be drawn.
\begin{itemize}
\item
The ODE (\ref{2-9}) has a 10 dimensional symmetry group of contact transformations $G$. The $O \Delta S$ (\ref{e3-14}) is only invariant under a 7-dimensional subgroup of contact transformations $G_0 \subset G$. The equation $p^{(3)}_{n+3}=0$ is actually invariant under the larger group $G$, but the lattice condition $h_{n+2}=c h_{n+1}$ is only invariant under $G_0$.

The situation is similar for the second order equation $y''=0$. The ODE is invariant under the group $SL(3, \mathbb{R})$ of point transformations. The corresponding $O \Delta S$ is invariant under a 6-dimensional subgroup of $SL(3, \mathbb{R})$ isomorphic to the group of general affine transformations of $\mathbb{R}^2$ \cite{DKW2000}.
\item
The Lie algebras of point symmetries of the $O \Delta S$ and the ODE are realized by identical vector fields (with the correspondence $(x,y) \leftrightarrow (x_n, y_n)$. The contact symmetry is however different: the derivative $p=u'$ is not replaced by $p^{(1)}_{n+1}$ but by $p^{(1)}_{n+1} - \frac{1}{2} h_n p^{(2)}_{n+2}$ (see (\ref{e3-16}) as opposed to $X_7$ in (\ref{2-11})).
\item
The contact transformation $X_7$ for the $O \Delta S$ (\ref{e3-14}) involves ``second order contact'', i.e. $\xi_n$ and $\phi_n$ in (\ref{e3-16}) depend on $p^{(1)}_{n+1}$ and $p^{(2)}_{n+2}$. In the continuous limit $h_{n+1} \rightarrow 0$ this reduces to first order contact ($p=u_x$ only). This is in agreement with B{\"a}cklund's theorem stating that contact transformations for an ODE are of most of order 1 \cite{Backlund}.
\end{itemize}

In \cite{AKO93} the authors introduced the concept of ``internal'' and ``external'' symmetries. External symmetries are defined on the entire jet space, internal ones only on the submanifold of the solutions of the equation. External symmetries refer only to strong invariants, internal symmetries also to weak ones. As stressed in \cite{AKO93} B{\"a}cklund's theorem \cite{Backlund} actually only applies to external symmetries.

In \cite{LTW1} we have shown that the only external symmetries of an $ O \Delta S$ are point ones. Here we have shown that the $O \Delta S$  can allow a class of higher symmetries that reduces to contact ones in the continuous limit. In the terminology of Ref \cite{AKO93} these are internal symmetries. They do not necessarily depend only on first order discrete derivatives.

It remains to determine whether this class of higher symmetries is actually useful, in particular whether it can be used to obtain solutions of ordinary difference schemes. To answer this question we are planning to study symmetry preserving discretizations of nontrivial ODEs that allow symmetry groups of genuine contact transformations \cite{Wafo}.

\section*{Acknowledgement}
We thank Vladimir Dorodnitsyn and Martin Thoma for helpful discussions. The research of PW was partly supported by a research grant from NSERC of Canada. LD and SC have been partly supported by the Italian Ministry of Education and Research, PRIN "Continuous and discrete nonlinear integrable evolutions: from water waves
to symplectic maps" from 2010. ZT thanks CRM, where parts of the research were carried out, for hospitality.


\end{document}